# EXCITATION OF THE AROMATIC INFRARED EMISSION BANDS: CHEMICAL ENERGY IN HYDROGENATED AMORPHOUS CARBON PARTICLES?

W. W. Duley

Department of Physics and Astronomy, University of Waterloo, Waterloo, Ontario, Canada, N2L 3G1

D. A. Williams

Department of Physics and Astronomy, University College London, Gower Street, London WC1E 6BT, UK

Email: wwduley@uwaterloo.ca

2## ABSTRACT

We outline a model for the heating of hydrogenated amorphous (HAC) dust via the release of stored chemical energy and show that this energy (~12 kJ/mole) is sufficient to heat dust grains of classical size (50-1000 Å) to temperatures at which they can emit at 3.3 μm and other "UIR" wavelengths. Using laboratory data, we show that this heating process is consistent with a concentration of a few percent of dangling bonds in HAC and may be initiated by the recombination of trapped H atoms. We suggest that the release of chemical energy from dust represents an additional source of excitation for the UIR bands relaxing the previous requirement that only stochastically heated molecules having fewer than ~ 50 atoms can produce emission at 3.3 μm.

Subject headings: ISM: lines and bands — ISM: molecules — infrared:ISM

1. INTRODUCTION

The ubiquitous infrared emission bands, now detected in a wide range of interstellar, circumstellar and extra-galactic sources, were first discovered by Gillett, Forrest and Merrill (1973). It was soon noted that the appearance of features at 3.3 and 11.3 µm in these spectra are indicative of aromatic CH groups as found in polycyclic aromatic hydrocarbons (PAH) such as coronene (Duley & Williams 1981) (DW 81). Emission at 3.4 µm, together with absorption at this wavelength in the diffuse interstellar medium (DISM), suggests that this PAH-like ($sp^2$ bonded) component is also mixed with aliphatic hydrocarbons ($sp^3$ bonded carbons). In DW 81, the relative proportion of $sp^2$ and $sp^3$ bonded carbons was postulated to depend on ambient conditions with the $sp^2$ component dominating in regions of high excitation, while $sp^3$ bonded hydrocarbons were suggested as the preferred composition for carbons in the DISM. Mixed $sp^2/sp^3$ bonded carbon compositions of this type are commonly found in hydrogenated amorphous carbons (HAC or a-C:H) (Angus & Hayman 1988). IR spectra of these materials in thin film and nanoparticle form measured in the laboratory provide good fits to interstellar absorption (Scott & Duley 1996, Dartois et al. 2007) and emission (Scott et al. 1997, Grishko & Duley 2000, Hu et al. 2006) spectra.

Understanding the mechanism responsible for the excitation of IR emission from these particles under interstellar conditions is key to determining the composition and structure of these emitters. DW81 attributed the IR bands to thermal emission by small carbon particles in a model that envisaged equilibrium excitation. While this may be appropriate for particles in circumstellar shells and in the envelopes of post AGB stars, the detection of emission in the DISM and in nebulae far from illuminating stars is incompatible with this mechanism. An alternative model, involving non-equilibrium emission from carbon nanoparticles stochastically heated to high temperatures by the absorption of individual photons from the interstellar radiation field was proposed by Sellgren (1984). The observed excitation temperature for emission at 3.3 μm was found to be consistent with the presence of emitters containing ~ 50 atoms. Given the spectral features detected, with bands at 3.3, 3.4, 6.2, 7.7, 8.6 and 11.3 μm, Leger & Puget (1984) and Allamandola et al. (1985) then proposed that such particles could best be described as large PAH molecules rather than as small solid particles having mixed $sp^2/sp^3$ composition. Since that time, sensitive measurements have failed to detect any *specific* PAH molecule in interstellar sources (Gredel et al. 2011, Searles et al. 2011, Galazudinov et al. 2010, Pilleri et al. 2009) even at abundance levels much less than those expected from chemical models, although the napthalene and anthracene cations may have been detected in one source (Iglesias-Groth et al. 2010). In addition, emission can still be detected in regions having little UV radiation (Uchida et al. 1998) implying that high energy UV photons are not required for excitation. Despite these limitations, the PAH model has come to be widely accepted by astronomers and "free-flying" PAHs are now presumed to constitute a major reservoir for carbon in the universe, with important consequences for heating, charge balance and chemistry in interstellar clouds.

In this paper we re-examine the primary assumption inherent in the "PAH hypothesis" (Tielens 2011) that only particles of molecular size can be responsible for emission at 3.3 μm and longer wavelengths. We find that this requirement can be relaxed if the emitters of the 3.3 μm and other IR bands are heated by the chemical energy released from reactions within interstellar grains. This model then postulates that the non-equilibrium emission of IR radiation can also be associated with larger carbon particles of mixed $sp^2/sp^3$ carbon composition.

## 2. TRANSIENT HEATING OF DUST GRAINS

The stochastic nature of the temperature of small interstellar dust grains in equilibrium with the interstellar radiation field has been discussed by Duley (1973), Greenberg & Hong (1974) and Purcell (1976). Time dependent excursions of grain temperature above the long-term average temperature are driven by the absorption of individual photons. This effect is most pronounced in tiny particles since the grain heat capacity is so small that the energy in a visible or UV photon is comparable to that contained in the grain. Re-emission of absorbed energy in the IR can then produce excesses at longer wavelengths (Duley 1974). As the amplitude of temperature excursions decreases rapidly with increasing particle size, it is evident that only very small particles (r ≤ 10 Å) can reach temperatures near 1000K after absorbing a UV photon. The detection of excess emission at 1- 3 μm in reflection nebulae (Sellgren et al. 1983) then led to Sellgren's (1984) suggestion that very small particles subject to temperature excursions were present in these objects. The equilibrium temperature for classical (ie. larger) dust grains in these objects was found to be insufficient to account for the observed excesses.

Regardless of the excitation mechanism, thermal emission at an elevated temperature, T(K), requires that the particle acquire a total energy

$$E(T) = n_p \int_{T_0}^{T} C(T) dT$$

where C is the grain heat capacity (J/mole K), $T_0$ is the initial grain temperature and $n_p$ is the number of moles of material in the particle. C(T) has been calculated for carbon particles having a mixed PAH/graphite composition by Draine & Li (2001) in a model that includes the effect of discrete low energy vibrational modes. Using their data to estimate E(T) we find that a particle containing 100 atoms ($n_p = 1.66 \times 10^{-22}$ mole) requires an excess energy of $\approx 2 \times 10^{-18}$ J (12.5 eV) to reach T = 1000K. As expected, it is evident that the absorption of individual photons can produce a temperature in this range in such small particles. However, a carbonaceous particle, typical of a small interstellar grain, with a volume V = $10^{-18}$ cm$^3$ and containing $\approx 10^5$ atoms ($n_p = 6 \times 10^{-19}$ mole) and would require $2 \times 10^{-15}$ J = $12.5 \times 10^3$ eV to reach the same T. This is obviously much too large an energy to be obtained from the interstellar radiation field. On a chemical scale, however, the energy required is still only $\approx$ 12 kJ/mole, considerably less than typical chemical bond energies of 350-800 kJ/mole.

Calculation of E(T) shows that interstellar grains of canonical size (50-1000Å) can emit at ~ 1000K if stored chemical energy of $\approx$ 12 kJ/mole, amounting to several percent of the total bond energy contained within the grain, is suddenly released. The trigger for a reaction of this kind would be an abrupt rise in the grain temperature to a point where the activation barrier for energetic reactions can occur (Greenberg 1976). A series of laboratory experiments documenting the explosive recombination of carbyne molecules to form amorphous carbon provides some insight into how this process might occur in practice (Wakabayashi et al. 2004, Yamaguchi & Wakabayashi 2004). In these experiments, carbon molecules obtained from the laser ablation of graphite, were condensed in an excess of inert gas on a surface at 6K. Warming this mixture to ~ 40K resulted in "violent energy release" with the emission of visible and near IR radiation from the sample. This highly exothermic reaction was attributed to the conversion of sp-bonded carbon molecules (acetylenic and cumulenic species) into sp$^2$-bonded compounds as found in amorphous carbon. The energy released in this conversion was found to be 0.2-0.4 eV/atom (19-38 kJ/mole), somewhat larger than the 12kJ/mole needed to increase the temperature of a large isolated particle to 1000K. Similar exothermic reactions have also been reported in the formation of $C_{60}$ and other fullerenes from small carbon clusters (Suzuki et al. 2001).

These experiments suggest that reactions within hydrogenated amorphous carbon grains could provide the chemical energy necessary to produce temperature excursions and emission in the near IR and longer wavelengths. In this context, it has been known for some time that HAC can contain up to ~ 60 atomic % of hydrogen. Much of this is tied up in a variety of hydrocarbon structures, but it has been estimated that 0.3-0.5 of this hydrogen is not chemically bound to carbon and is instead held in the form of $H_2$ and H atoms trapped at interstitial and weakly bonded internal sites with binding energies of 0.05-0.2 eV (Sugai et al. 1989). These H atoms would accumulate in HAC under the low T conditions prevalent in interstellar clouds, and represent stored chemical energy as the exothermicity of the H + H → $H_2$ reaction is 436 kJ/mole. With the assumption that only H + H recombination reactions would be responsible for transient heating to T = 1000K, the required concentration of trapped H prior to the thermal event is $\eta = 2(12/436) = 0.055$.

A more realistic picture is one in which HAC grains accumulate dangling bonds as well as trapped H atoms on exposure to H atoms and other reactive species in interstellar clouds. This effect has been well documented in laboratory studies (Biener et al. 1994, Jariwala et al. 2009) when amorphous carbon is exposed to a flux of H atoms. Hydrogenation reactions are highly exothermic and occur with little activation energy, but they can also produce radical species with dangling bonds which would add to the stored chemical energy. Recombination of these fragments involves the re-arrangement of C−C, C=C and C≡C bonds analogous to the consolidation type of reactions observed by Wakabayashi et al. (2004). Thermal spiking in HAC grains is then postulated to involve a process in which H atoms and other reactive species accumulate until they reach a concentration of several percent at which point they participate in a thermal runaway type of reaction. Photodissociation of molecules following absorption of photons from the interstellar radiation field is also expected to contribute to the overall concentration of radical species internal to HAC.

## 3. DISCUSSION

Thermal emission spectra (Scott et al. 1997, Grishko & Duley 2002) as well as surface enhanced Raman spectra (Hu et al. 2006, Hu & Duley 2008 a,b) of HAC and related carbon nanoparticle solids are found to be generally consistent with observed interstellar and circumstellar IR emission spectra in a variety of sources. Indeed, the agreement between laboratory and observed spectra (Fig. 1) extends to include reproduction of both the profile and the width of the 3.3 μm and other emission bands (Hu & Duley 2007, 2008 a,b). In addition, changes in the composition of these solids have been found to yield spectra that are consistent with spatial variations within emitting objects (Hu et al. 2008a) and with the distinction between types A, B and C spectral classes as defined by Peeters et al. 2002. To our knowledge, there are no comparable fits to astronomical spectra using laboratory data for gas phase PAH.

Empirical data is then supportive of a model in which non-equilibrium IR emission occurs from small solid particles with a composition similar to that of HAC dust as produced in numerous laboratory simulations (Scott et al. 1997, Hu et al. 2006, Dartois et al. 2005, 2007,

Jager et al. 2011). This material contains both aromatic and aliphatic hydrocarbon species and is a random network of sp, $sp^2$ and $sp^3$-bonded carbons. We suggest that much of the energy produced in exothermic chemical reactions initiated by H + H recombination at internal trapping sites heats the grain to temperatures where it emits in the near IR. The spectrum of this emission would be characteristic of the chemical composition of HAC which includes PAH, aliphatic and fullerene/fullerane structures with the relative strength of emission features from each chemical component dependent on the overall chemical and radiative history of the grain material. In equilibrium in interstellar sources, this would be a cyclic process in which the H atom concentration builds up to a point where the thermal event occurs. Newly formed dust would be expected to consolidate in response to this transient thermal heating with aliphatic structures combining to form rings and other compact carbon structures. Eventually, a stable composition would be reached in which the material evolves into an extended network of aromatic rings connected by aliphatic chains. This implies that UIR emission should be strong in H-rich regions where dust has been newly formed and weaker in regions where H has been converted to $H_2$ and the dust is more evolved. Thermal heating via chemical reactions internal to HAC may also contribute to the excitation of IR emission from fullerenes and would explain why fullerenes have now been detected in a variety of H-rich circumstellar environments (Garcia-Hernandez et al. 2011, 2010). The hydrogenation of fullerenes during such a process would potentially form fulleranes resulting in IR emission at 3.4 μm (Linnolahti et al. 2006). It is interesting that these molecules do not emit at 3.3 μm because they lack $sp^2$-bonded CH groups.

A side effect of this chemical heating involves the sudden release of warm $H_2$ molecules trapped inside the HAC solid which may account for the good correlation frequently observed between the spatial distribution of 2.18 μm $H_2$ (v = 1 – 0, S (1)) filaments and 3.3 μm emission in a number of sources (Rouan et al. 1999, An & Sellgren 2003). This model predicts that excited $H_2$ emanating from HAC would consist of two components: one from freshly generated $H_2$ with an excitation characteristic of the formation process, and another involving the release of adsorbed $H_2$ molecules. The excitation of the latter should be in thermal equilibrium with the grain temperature T ≈ 1000K.

A related scenario involving the possible role of chemical energy in the crystallization of interstellar silicate grains has recently been proposed by Kaito et al. (2007) and Tanaka et al. (2010). In their model, heat liberated from the crystallization of an amorphous carbon on the surface of silicate dust heats an underlying silicate core to a temperature (~ 1000K) allowing it to convert to crystalline form. They estimate that this would require ~1-10% of the material to consist of reactive molecules. This is similar to our estimate of the concentration of adsorbed H needed to produce thermal 3.3 μm emission from HAC, suggesting that the two processes may be related and that stored chemical energy could play an important part in determining the overall radiative properties of interstellar dust. For example, while it is customary to assume that continuum emission from dust at long wavelengths arises from thermal heating of dust by absorption of short wavelength photons, the release of chemical energy from dust would augment this process.

This research was supported by a grant from the NSERC of Canada

Figure caption

1. Comparison between the 800K emission spectrum of HAC and Type A and Type B interstellar spectra (Peeters et al. 2002). The laboratory spectrum contains an enhanced component near 3075 cm$^{-1}$ (3.25 μm) attributable to olefinic CH groups (Hu & Duley 2008b). This feature may also be present at a reduced level in the interstellar spectra. HAC was produced by vacuum deposition as described previously (Grishko & Duley 2000). The emission spectrum was measured in vacuum with a Bomem MB-100 FTIR spectrometer.

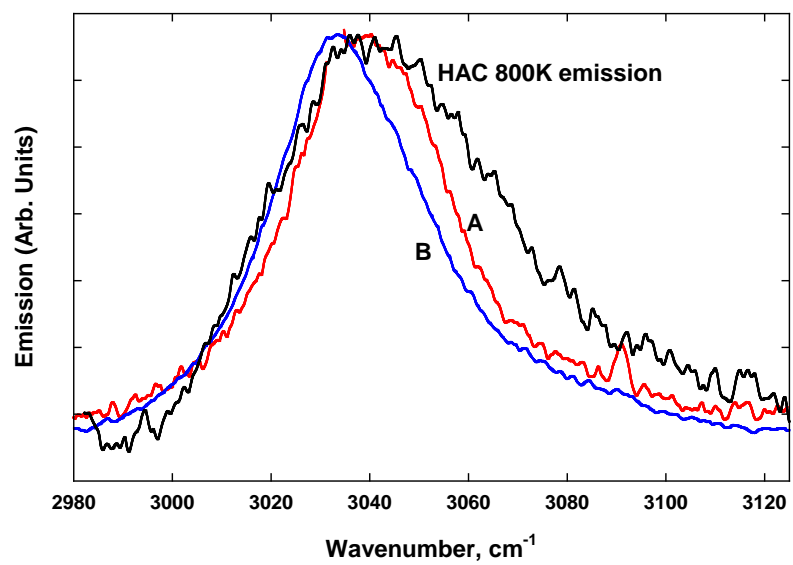

Fig. 1